\documentclass[conference]{IEEEtran}
\IEEEoverridecommandlockouts
\usepackage{cite}
\usepackage{amsmath,amssymb,amsfonts}
\usepackage{algorithmic}
\usepackage{graphicx}
\usepackage{textcomp}
\usepackage{xcolor}
\usepackage{subfigure}
\usepackage{hyperref}
\usepackage[T1]{fontenc}

\newtheorem{definition}{Definition}[section]

\def\BibTeX{{\rm B\kern-.05em{\sc i\kern-.025em b}\kern-.08em
    T\kern-.1667em\lower.7ex\hbox{E}\kern-.125emX}}
\begin{document}

\title{An attack resilient policy on the tip pool for DAG-based distributed ledgers
\thanks{This work is funded by the IOTA Foundation.} 
}

\author{\IEEEauthorblockN{Lianna Zhao\IEEEauthorrefmark{1},
Andrew Cullen\IEEEauthorrefmark{2},
Sebastian Mueller\IEEEauthorrefmark{3}, 
Olivia Saa\IEEEauthorrefmark{2}, and
Robert Shorten\IEEEauthorrefmark{1}}
\IEEEauthorblockA{\IEEEauthorrefmark{1} Dyson School of Design Engineering, Imperial College London, London, U.K. (l.zhao20@imperial.ac.uk)}
\IEEEauthorblockA{\IEEEauthorrefmark{2} IOTA Foundation, 10405 Berlin, Germany.}
\IEEEauthorblockA{\IEEEauthorrefmark{3} Aix Marseille Universite, CNRS, Centrale Marseille, I2M - UMR 7373, 13453 Marseille, France.}
\thanks{}}

\maketitle
\thispagestyle{plain}
\pagestyle{plain}

\begin{abstract}
 This paper discusses congestion control and inconsistency problems in DAG-based distributed ledgers and proposes an additional filter to mitigate these issues. Unlike traditional blockchains, DAG-based DLTs use a directed acyclic graph structure to organize transactions, allowing higher scalability and efficiency. However, this also introduces challenges in controlling the rate at which blocks are added to the network and preventing the influence of spam attacks. To address these challenges, we propose a filter to limit the tip pool size and to avoid referencing old blocks. Furthermore, we present experimental results to demonstrate the effectiveness of this filter in reducing the negative impacts of various attacks. Our approach offers a lightweight and efficient solution for managing the flow of blocks in DAG-based DLTs, which can enhance the consistency and reliability of these systems.
\end{abstract}


\section{Introduction}\label{def: Intro}
Distributed ledger technologies (DLTs) have gained significant attention for their potential to revolutionize how we store, manage and transfer digital assets. DLTs, such as blockchains, have traditionally relied on a linear chain of blocks. However, this approach has limited scalability and efficiency, particularly when faced with high volumes of blocks. Directed Acyclic Graph (DAG)-based DLTs offer an alternative solution to this problem by using a DAG structure to organize blocks. This more flexible structure allows for higher scalability and lower latency, as each block can reference multiple previous blocks rather than relying on a linear chain.

One of the benefits of DAG-based DLTs is the less restrictive writing access enabled by their DAG structure. Ideally, this can make the mempool redundant and allow for a more efficient design of the block dissemination. However, this also introduces challenges in controlling the rate at which blocks are added to the ledger and preventing spam attacks. Writing access can be controlled by Proof of Work (PoW), e.g.~\cite{PHANTOM, DAGKNIGHT}, Proof-of-Stake (PoS)-based lottery, or permissioned setup, e.g.~\cite{Gagol2019, Bullshark, CordialMiners}.
We will focus our attention in this work on a specific DAG-based DLT architecture, based on IOTA~\cite{popov2020coordicide,muller2022tangle}. This architecture does not restrict writing access to special validator nodes but enables all nodes to participate through congestion control on a different layer. This congestion control mechanism must be resilient against malicious actors who wish to compromise the throughput of honest actors by spamming the network with their blocks. To address this challenge, we propose a filter, in addition to the scheduler and drop-head policy proposed in \cite{cullen2021access}, which prevents referencing of old blocks. This mitigation strategy corresponds to the two following components that seem vulnerable to liveness and consistency attacks. 

First, the proposed protocol uses a congestion control algorithm on the underlying P2P layer. Each honest node adjusts its issuing rate using an AIMD rate setter; for more details, see \cite{cullen2021access}\cite{zhao2021secure}. Every node keeps a separate queue for each issuer inside the scheduler that regulates the gossiping of the blocks. The queues of the malicious or faulty nodes that exceed their quota of blocks, will blow up (in the absence of a mitigation strategy) and eventually lead to inconsistencies in the ledger. 

Second, in the proposed protocol, the \emph{tip selection algorithm} (TSA), which determines how new blocks are attached to the existing DAG, plays a crucial role. Each time a node issues a new block, it \emph{approves} previous existing blocks from the DAG which have not yet received any approvals. These unapproved blocks are called tips. The set of tips eligible for selection is called the tip pool, and the size of this set is a critical metric of the DAG-based DLTs, e.g.~\cite{penzkofer2020parasite,cullen2019distributed,he2019securing} \cite{zhao2021secure}\cite{stabilityOfTips}. The situation where the tip pool size becomes excessively large due to faulty or malicious behaviour is considered the most problematic, as it eventually leads to liveness issues and inconsistencies. 
\begin{figure*}[ht]
\centering\includegraphics[width=1.7\columnwidth]{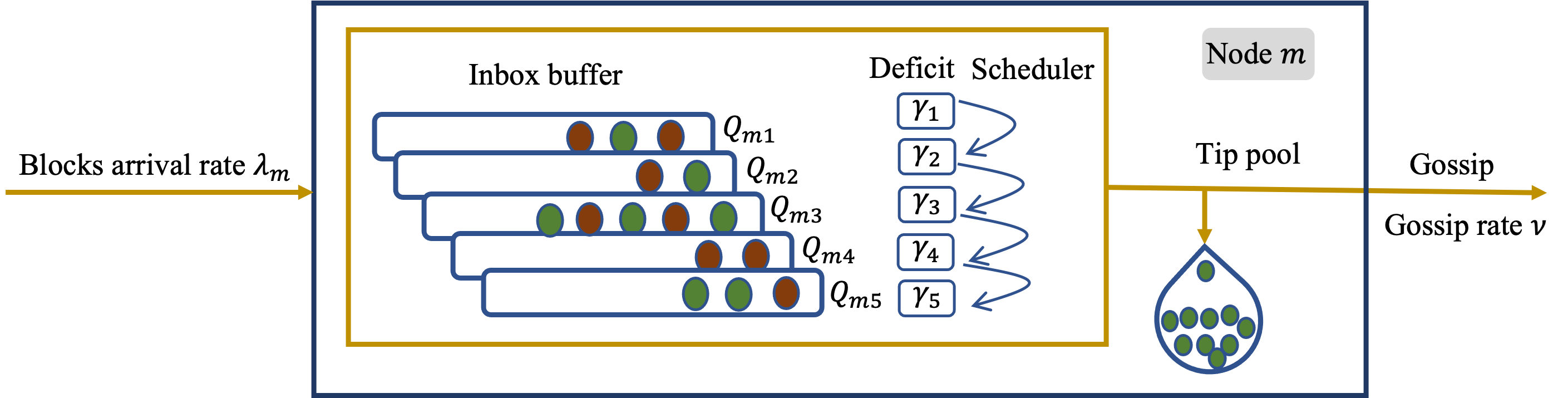}
\caption{Data flow for node $m$ in the network.} 
\label{fig: Network}
\end{figure*}

Our main focus of this paper is mitigating spamming attack scenarios and inconsistency problems in tip selection. Spamming attackers send a large number of blocks into the network and inflate the buffer size of honest nodes. The consequence of this attack is causing large delay of blocks issued by honest nodes, resulting in a large tip pool size and affecting the liveness of the network. We will also consider spammers that send different streams of blocks to each of their neighbors, causing inconsistency issues. The TSA is a key component of the protocol for mitigating the negative effects of these attacks by ensuring honest nodes attach their blocks to the correct part of the DAG.

The goal of this paper is to design a secure TSA for DAG-based ledgers to provide resilience against tip pool inflation and inconsistency problems caused by spammers. To reduce the network burden, we adopt the buffer management component proposed in \cite{cullen2021access}, also known as drop-head policy, as the first filter to deal with spamming blocks. Furthermore, our solution for the tip pool regulation (named the \emph{past-cone confirmation time (PCT)} condition) checks the validity of blocks after being scheduled, effectively regulating the size of the tip pool and ensuring the consistency of the ledger. Specifically, the PCT condition is introduced to manage the inconsistency problem caused by the previous tip selection and limits the size of the tip pool, especially in the presence of attacks.

\section{Related work}
Although the design of congestion control algorithms has been a widely studied topic in the domain of computer networking, when it comes to distributed ledgers, this topic is relatively new, and developing access control algorithms in this specific context poses unique and unprecedented challenges. This is primarily due to the competitive and hostile nature of the environments in which ledgers are designed to operate. A novel design paradigm for DAG-based DLTs is introduced in \cite{cullen2021access} to enable the integration of reputation-based access control
for the first time. \cite{cullen2021access} presents
a rate-setting algorithm, a buffer management component, and a scheduler, which enables the system to be resilient against any malicious agents attempting to acquire a greater portion of network resources than their fair allocation. An enhanced access control algorithm specifically for DAG-based IOTA Tangle is proposed in \cite{zhao2021secure} to improve the network's security and resilience. Specifically, to address issues related to spamming and multi-rate malicious attackers, a blacklisting algorithm that employs a reputation-weighted threshold is introduced.

Tip selection has been a central part of the IOTA protocol since its inception, and numerous prior works have studied TSAs and their implications.
The authors in \cite{muller2022tangle} show under reasonable assumptions that a growing tip pool size grows results in larger confirmation times. The authors in \cite{penzkofer2021impact} presented an analysis of processing two classes of blocks and the resulting delay time and further proposed a more general model for predicting tip pool size. In \cite{ferraro2022feedback}, to tackle the tip inflation attack, the author provided a fluid model to analyse the attack and designed a feedback control to regulate the number of tips that should be selected. 
According to the papers \cite{penzkofer2021impact, popov2018tangle, cullen2019distributed}, the average number of tips in each node's tip pool is approximately two times the product of the average time delay of blocks and the block arrival rate. Here, the time delay is measured from when blocks are issued to when blocks arrive at the tip pool, while the delay is averaged over all blocks issued by all nodes. 
A first mitigation strategy for tip pool attacks was recently studied in \cite{Cam:23}, which limits the size of the local tip pools and drops the oldest tips if the maximal tip pool size is achieved. 

In our work, we propose a more sophisticated version that also allows controlling to which parts of the DAG new blocks are attached. This feature is crucial for the construction of finality gadgets and to prevent ``parasite-chain'' attacks as discussed in \cite{penzkofer2020parasite}.

\section{System model}
We begin by describing the relevant components of the data flow of a node, $m$, as depicted in Figure \ref{fig: Network}.

 According to the network's congestion condition, as indicated by buffer lengths, each node adjusts its issuing rate using an AIMD rate setter (the interested reader can refer to \cite{cullen2021access} \cite{zhao2021secure}).
In node $m$'s inbox buffer, the inbox is split into $N$ queues to identify blocks issued by different nodes. For example, for blocks issued by node $i$, its blocks are assigned to $Q_{m_i}$ in node $m$'s buffer. The Deficit Round Robin (DRR)- scheduler is used for scheduling the next message to be forwarded and added to the tip pool (the interested reader can refer to \cite{cullen2021access} \cite{shreedhar1995efficient}). Given this background, we introduce four possible states of a node in the network \cite{cullen2021access} \cite{zhao2021secure}. We assume the issuing rate of nodes is defined as $\lambda_m$ and $\tilde{\lambda}_m = \frac{\nu \cdot rep_m}{\sum_{i \in \mathcal{N}}{rep_i}}$ is the guaranteed allowed rate, where $rep_m$ is a numeric reputation value which is associated to a node $m$ and $\nu$ is the scheduling rate of node $m$ \cite{cullen2021access,zhao2021secure}.

\begin{itemize}
\item[(1)] A node $m$ is said to be {\em inactive} if the issuing rate $\lambda_m=0$.
\item[(2)] A node $m$ is said to be {\em content} if it issues blocks that can be approximated as a Poisson process with a fixed rate parameter $\lambda_m \leq \tilde{\lambda}_m $.
\item[(3)] A node is said to be {\em best-effort} if it
issues at rate $\lambda_m > \tilde{\lambda}_m$ under the rate control policy imposed by the access control algorithm.
\item[(4)] A node is said to be {\em malicious} if it deviates from the designed protocol, including the rate setting algorithm and the forwarding algorithm \cite{cullen2021access,zhao2021secure}.
\begin{itemize}
\item For attackers deviating from the rate setting algorithm, they issue blocks at a rate far above the allowed protocol. We name this type of attacker spamming attackers.
\item For attackers deviating from the forwarding algorithm, they send a different stream of blocks to different neighbours while each stream obeys the rate control policy. We name this type of attacker multi-rate attackers.
\end{itemize} 
\end{itemize}

Normally, scheduled blocks are then added to the tip pool directly. But in the presence of attacks, such as spamming attacks, a large number of blocks are issued by a malicious node, and this burst of traffic inflates the node inbox. The consequence of this congestion is that the delay of each block increases, many old tips are added to the tip pool, and then the tip pool size keeps increasing. Finally, regarding the ledger, both the width of the DAG structure and the confirmation time (Refer to Definition~\ref{def: ConfirmedTX}) for blocks increase, affecting the confirmation efficiency of DAG-based DLTs. Further, as mentioned in Section~\ref{def: Intro}, the ledger might become inconsistent when some nodes pick up blocks that other nodes dropped.
Hence, we propose the PCT condition to protect against adversarial attacks and ensure the ledger's consistency. 
To facilitate exposition, we present some further notations and definitions that will be used in the remainder of the paper.

    \begin{figure}[ht]
    \centering
\includegraphics[width=0.9\columnwidth]{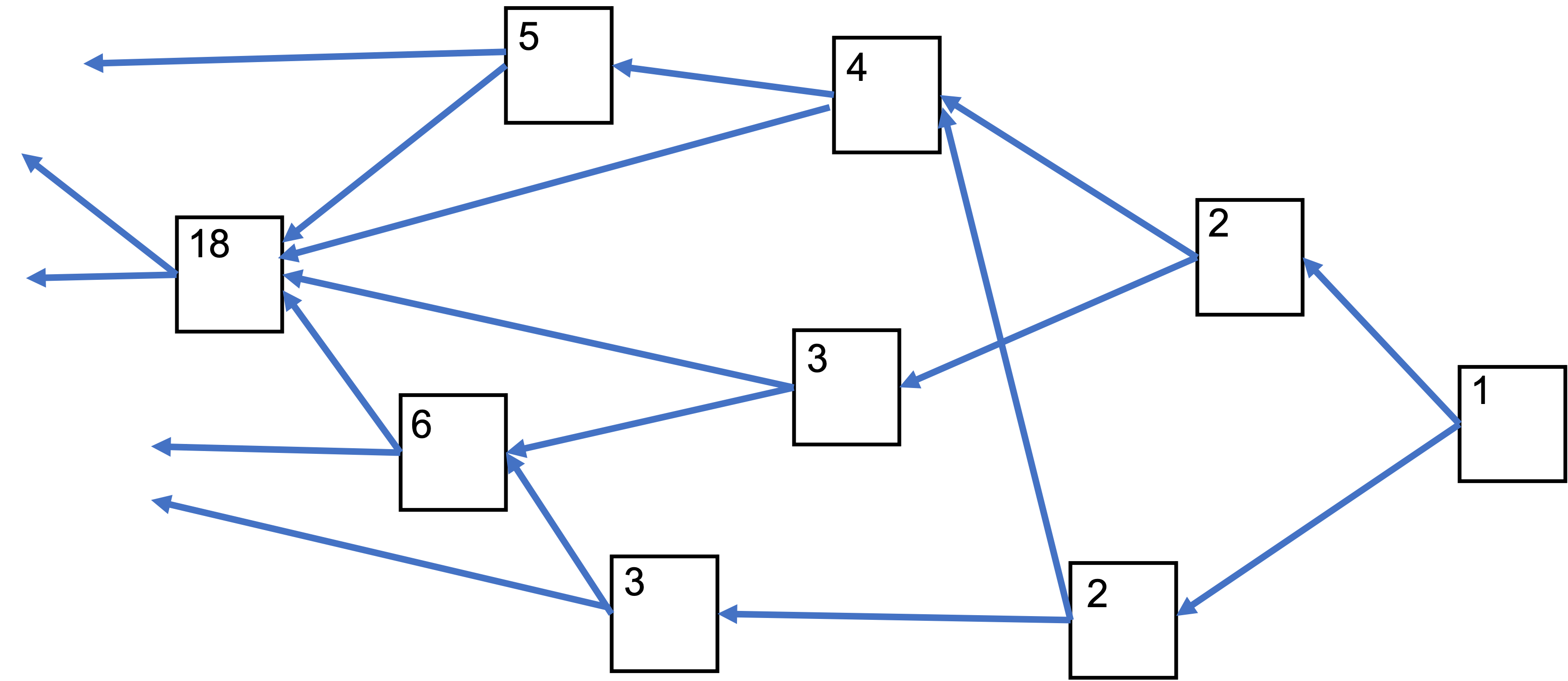}
    \caption{Cumulative weight representation (The number at the left top is the cumulative weight.)}
    \label{fig: Cw}
    \end{figure}

    \begin{figure*}
    \centering
\includegraphics[width=1.5\columnwidth]{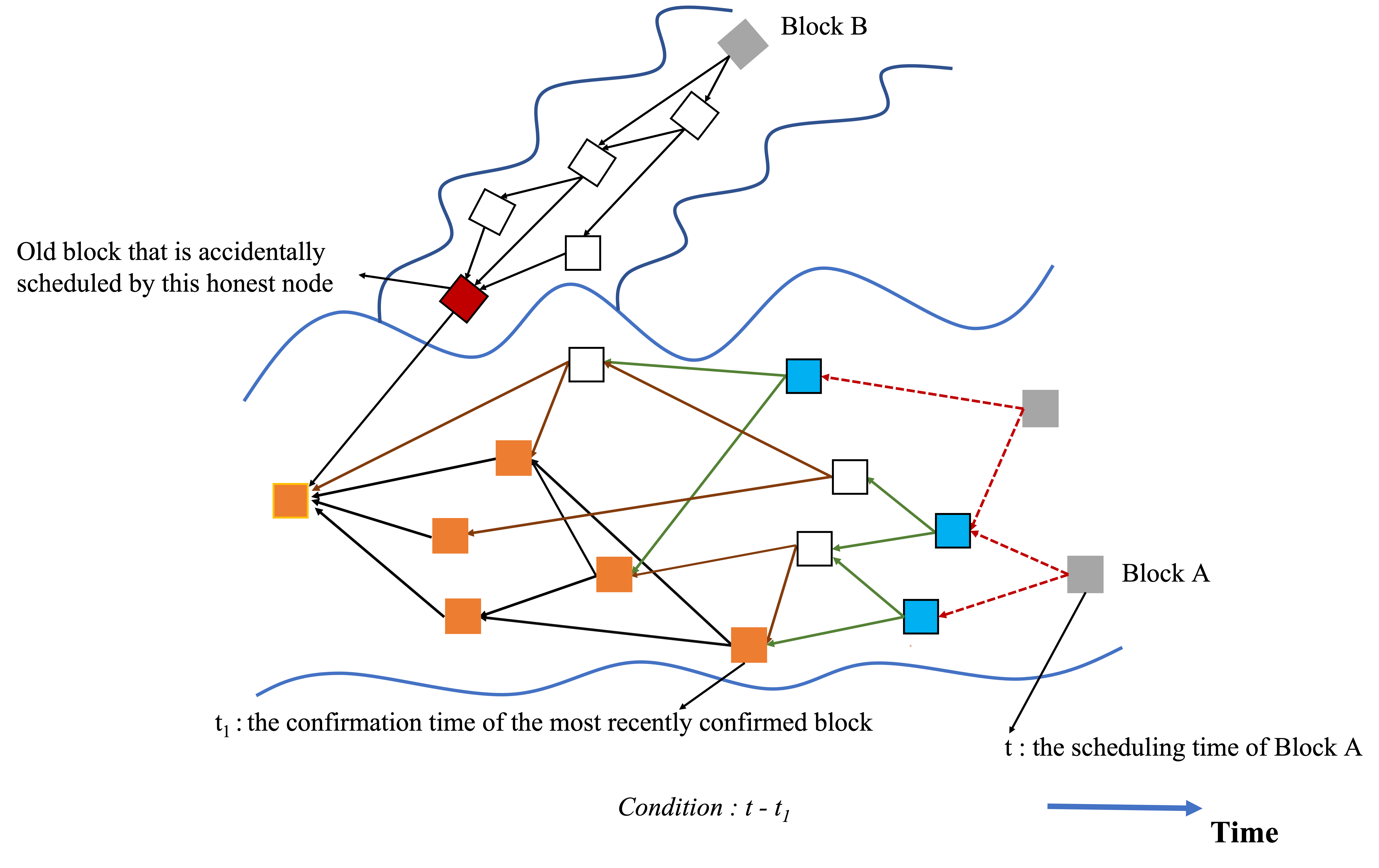}
    \caption{PCT condition.}
    \label{fig: BFS}
    \end{figure*}
    
\begin{definition}[Cumulative weight\cite{popov2018tangle}]
The cumulative weight of block $A$ is one plus the number of scheduled blocks that directly or indirectly approve block $A$. This is depicted in Figure \ref{fig: Cw}.
The number at the left top of a block is its cumulative weight. Since there might be discrepancies about which blocks the nodes see due to network delays, we sometimes refer to the local perception of node $m$ regarding the cumulative weight of a block (or the cumulative weight of a block in node $m$).
\end{definition}
 
 \begin{definition}[Disseminated block] A block is defined as being disseminated when all honest nodes in the network receive it.
\end{definition}

\begin{definition}[Dissemination rate \cite{cullen2021access}\cite{zhao2021secure}] We denote the rate of dissemination of all blocks issued by node $i$ as $DR_{i}$. The dissemination rate of all blocks is denoted $DR$ where $DR\triangleq \sum_{i=1}^ N DR_{i}$. 
\end{definition}

\begin{definition}[Scaled dissemination rate \cite{cullen2021access}\cite{zhao2021secure}]
The reputation-scaled dissemination rate is defined as the dissemination rate of this node divided by the node’s reputation value.
\end{definition}

\begin{definition}[Confirmed block\cite{popov2018tangle}]\label{def: ConfirmedTX}

If the cumulative weight of a block in node $m$, is larger than a preset cumulative weight threshold, we define this block as confirmed in node $m$. But only if the block is confirmed by all honest nodes in the network, we say this block is \emph{fully} confirmed \cite{cullen2023distributed}.
For example, after a block $A$ is scheduled, its weight is added to the cumulative weight named $CW_A$ of itself and to all blocks in this block $A$'s past-cone. If $CW_A$ is larger than a preset cumulative weight threshold $CW_T$ defined in Table \ref{tab: notation} in every network node, then we mark this block as fully confirmed.
\end{definition}

\begin{definition}[Confirmation rate]
The rate of blocks issued by node $i$ becoming \emph{fully} confirmed is denoted by $CR_i$ and $CR \triangleq \sum_{i=1}^ N CR_{i}$. 
\end{definition}

\begin{definition}[Scaled confirmation rate] 
The reputation-scaled confirmation rate of node $i$ is defined as the confirmation rate of node $i$ divided by the node’s reputation value, and it is calculated by $CR_{i}/rep_i$\footnote{We study this reputation-scaled metric because we want to ensure the fairness of the system. Here, to be specific, we want to ensure the max–min fairness of the confirmation rate. Specifically, the allocation is said to be max-min fair if the confirmation rate increase of node $i$ deceases another node $m$'s confirmation rate with equal or smaller reputation-scaled confirmation rate \cite{cullen2021access}.}. 
\end{definition} 

As previously mentioned, there are mainly two components used in this paper to deal with the spamming issue and the inconsistency problem. The first one is borrowed from \cite{cullen2021access}, in which a buffer management algorithm is used as a filter for spamming attacks. Namely, for node $m$, when the total buffer size is larger than the maximum buffer size $W_{Max}$, the eldest block issued by the node with the largest reputation-scaled queue length will be dropped. This process is executed repeatedly until the total buffer size is smaller than $W_{Max}$. The second one, which is the focus of this paper, is past-cone confirmation time (PCT) condition.

\subsection{PCT condition}
In order to regulate tip pool inflation and inconsistency, we introduce the PCT condition.

\begin{definition}[Past-cone confirmation time (PCT)] 
The PCT of a block is defined as the difference between the scheduling time of the block and the confirmation time of the most recently confirmed block in its past-cone.
\end{definition}

As illustrated in Figure \ref{fig: BFS}, the orange blocks represent confirmed blocks, while the white blocks denote unconfirmed ones. The grey blocks symbolize the newly scheduled blocks or tips, with one designated as block $A$ and another as block $B$. The PCT mechanism operates as follows: if the most recent confirmed block in the past-cone of the referenced block is significantly dated, the block is not included in the tip pool. Adhering to the PCT condition, a new incoming block is incorporated into the tip pool only if the time difference between block $A$'s scheduling time, denoted as $t$, and the confirmation time of the most recently confirmed block's time, $t_1$, within block $A$'s past-cone is less than the predetermined PCT condition threshold, $\mathrm{PCT_{th}}$.

Conversely, it may be feasible that a particular block (identified as the red block within the subbranch) could be scheduled by only a small subset of honest nodes, added to their tip pool, and subsequently selected as a tip. Such a block cannot accumulate sufficient weight quickly, and tips that approve the red block will eventually fail the PCT condition.

Calculation of a block's PCT requires a search for the most recent confirmed block within the incoming block's past-cone, which is conducted using a breadth-first search (BFS) algorithm. It is important to note that the search within this branch will be terminated once the issuing time of the block under investigation surpasses the predetermined maximum depth of BFS, denoted as $T_{Max_{i}}$.

\section{Simulations} \label{sec: simulations}
\begin{table}[ht]
\caption{Parameters and selected values for the simulation.}
\centering
 \begin{tabular}{c|l|l}
 \hline\hline
 Parameter  &Definition          &  Value (units)\\
 \hline
 $\mathrm{N}$                     & The total number of nodes  & 20\\
 $\mathrm{CW_T}$                    & Cumulative weight threshold        & 25\\
 $\mathrm{T_{Max_{i}}}$          & Maximum depth of BFS  & 80 (s)\\
  $\mathrm{PCT_{th}}$                & PCT threshold          &25 (s)\\
  $\mathrm{Max}_{\mathrm{Inbox}}$            & Maximum inbox threshold   & 200\\
  $\mathrm{\nu}$   & The scheduling rate of nodes  & 20 (blocks per second) \\
 \hline\hline
\end{tabular}
\label{tab: notation}
\end{table}

\begin{figure}[ht]
\centering
\includegraphics[width=0.95\columnwidth]{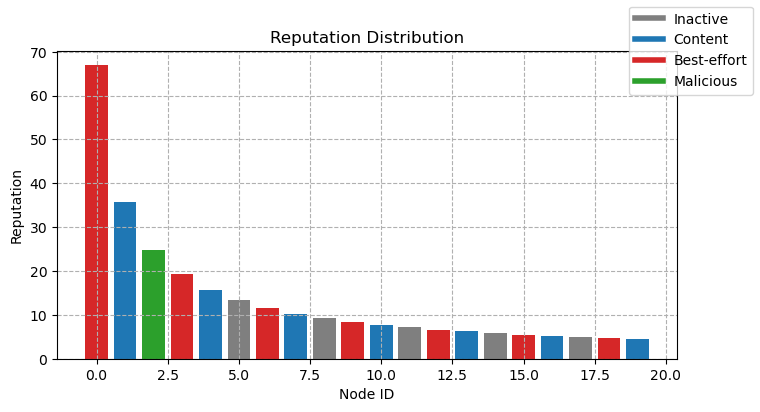}
\caption{Reputation distribution follows a Zipf distribution with exponent 0.9. As shown by each bar color, nodes are Best-effort in red, Content in blue, Inactive in gray, and Malicious in green \cite{zhao2021secure}.}
\label{fig: RepDist}
\end{figure}

The simulation\footnote{The code can be found in https://github.com/Mona566/TSC-condition.git} results are produced using a Python simulator. The parameters of the simulations, as outlined in Table~\ref{tab: notation}, are as follows. The number of nodes in the network is set to be $N=20$. Nodes are connected in a random 4-regular graph topology. 
The communication delay between each pair of nodes $i,j$ is uniformly distributed with values between $50$ms and $150ms$.
As shown in Figure \ref{fig: RepDist},
node reputations are assumed to follow a Zipf distribution (as measured from account balances in the IOTA network). The simulation results are averaged over 10 Monte Carlo simulations, each of which is 800 seconds. The PCT threshold is set as 25 seconds, and the cumulative confirmation weight threshold, $CW_T$, is set as 25 (the low threshold is propositional to slow issuing rate). The BFS condition we use for stopping searching is 80 seconds. The maximum inbox threshold is 200.

\textbf{Remark}: it is important to note that the specific performance figures indicated in these results are highly dependent on the parameters chosen, so they do not reflect actual DLT network performance. For example, by simply increasing the scheduling rate, one can immediately increase the confirmation rates and reduce the confirmation latencies. We focus instead on steady state behaviours and comparative results here.

Here, we consider the following scenarios.

\begin{itemize}
\item [$A_1$:] Attackers deviating from the rate setting algorithm, which are spamming attackers.

\item [$A_2$:] Attackers deviating from the forwarding algorithm, which are multi-rate attackers.

\item [$A_3$:] As a contrast experiment, we consider
the same attack scenario as $A_2$, but without implementing the PCT condition.

\end{itemize} 

\subsection{$A_1$: Single spamming attacker}
Here we consider the attack $A_1$, in which a spamming attacker is in the network.

\begin{figure*}
	\centering
    \subfigure[]{
    	\begin{minipage}[b]{0.3\linewidth}
   		\includegraphics[width=6.1cm, height=3.5cm]{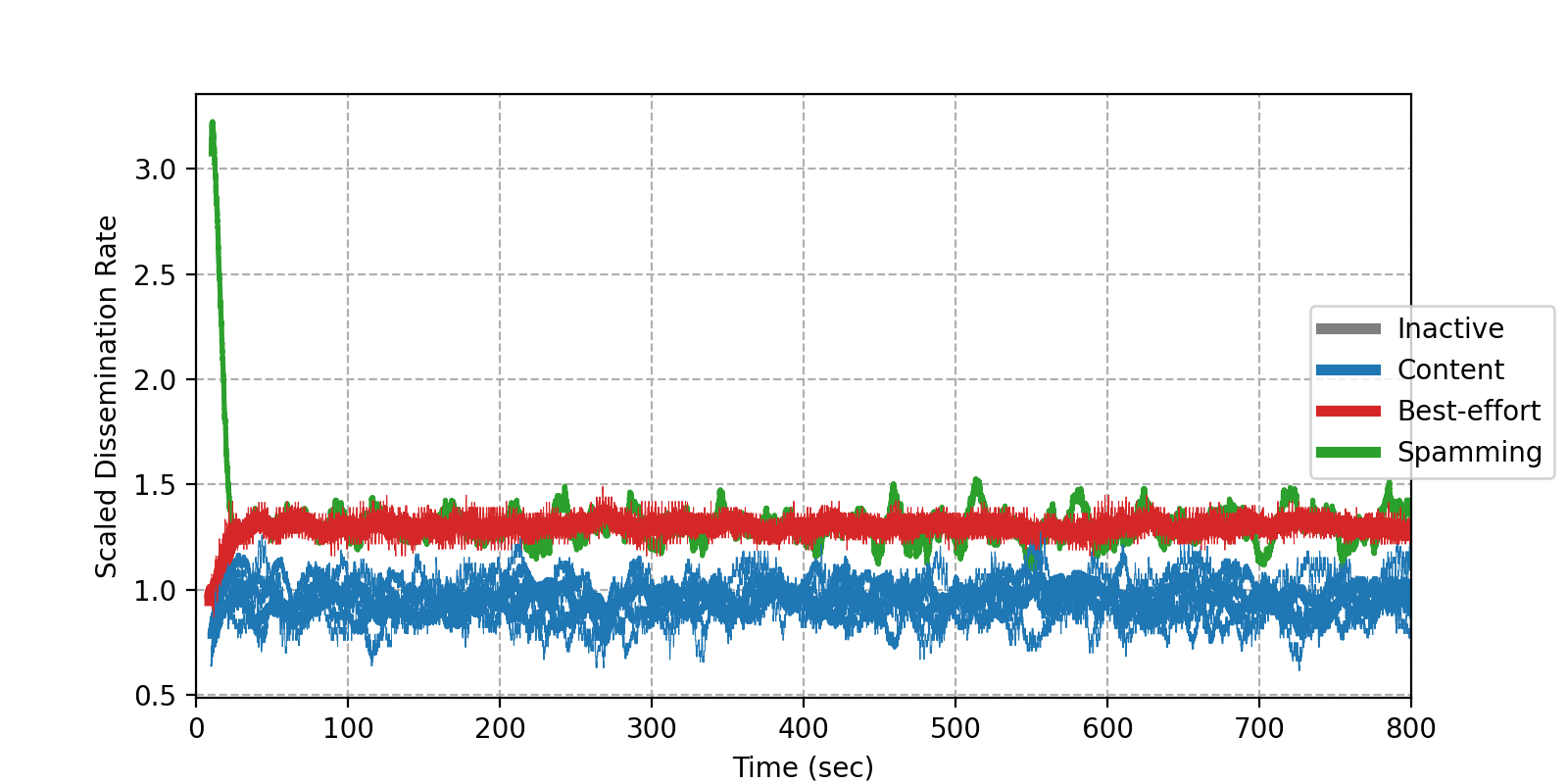}
    	\end{minipage}
	\label{fig:ScalSinSpaDisRate}
    }
    	\hspace{1mm}
    	\subfigure[]{
		\begin{minipage}[b]{0.3\linewidth}
			\includegraphics[width=6.1cm, height=3.5cm]{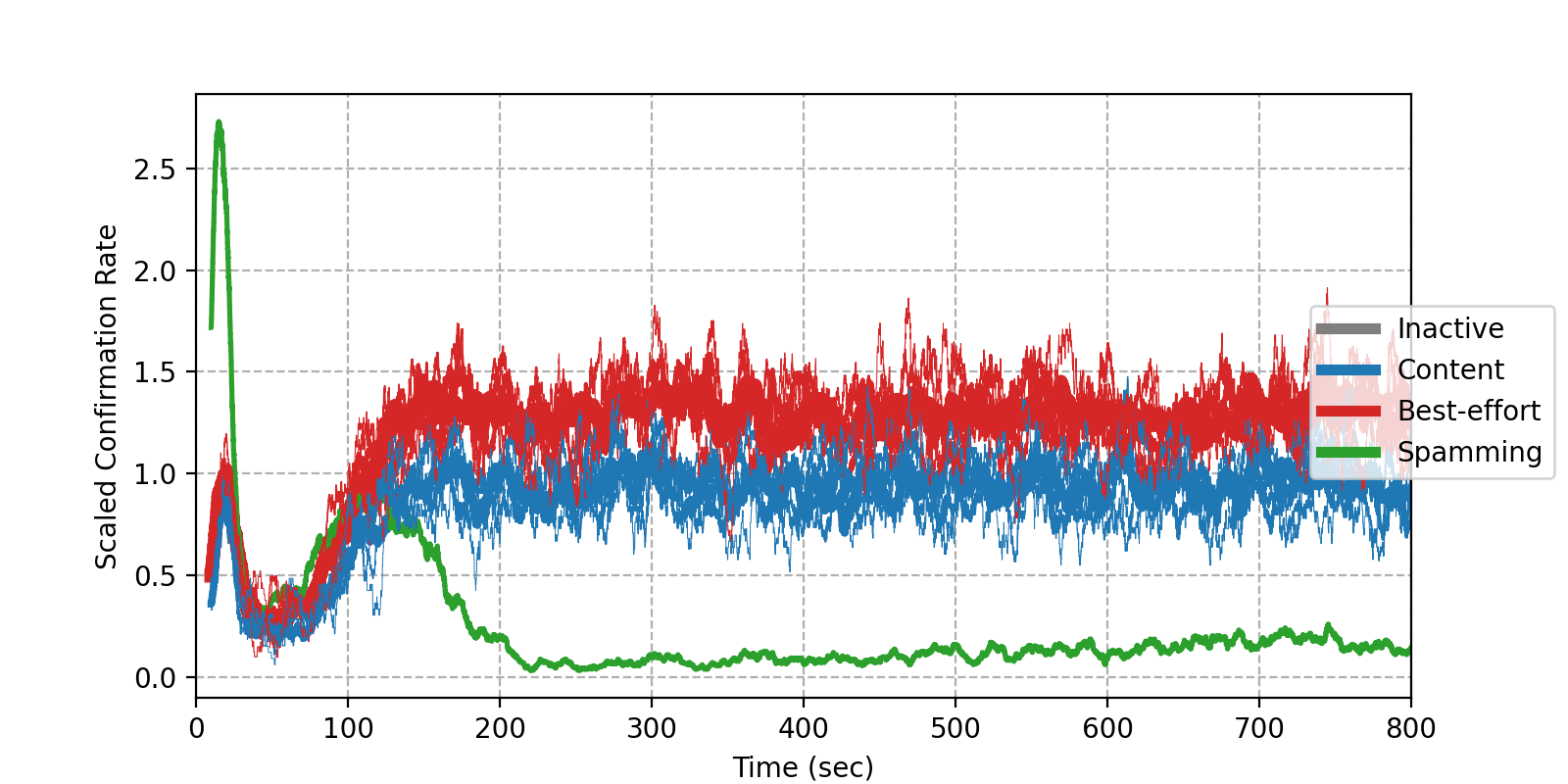}
		\end{minipage}
		\label{fig:ScaSinSpaConfirRate}
}
    	\hspace{1mm}
    	\subfigure[]{
		\begin{minipage}[b]{0.3\linewidth}
			\includegraphics[width=6.1cm, height=3.5cm]{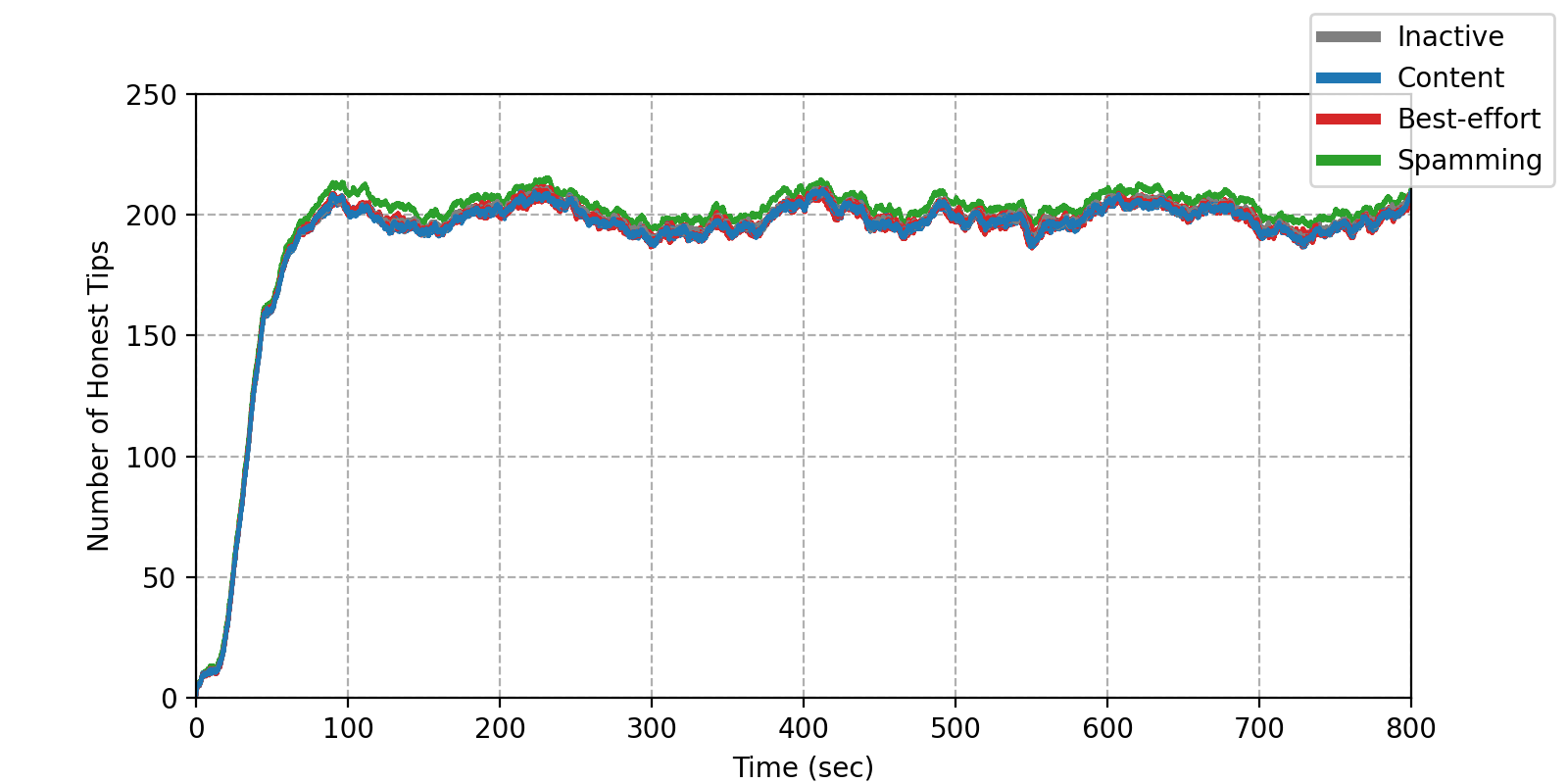}
		\end{minipage}
		\label{fig:Tip_size}
}
    	\hspace{1mm}
	\hspace{1mm}
    \subfigure[]{
    	\begin{minipage}[b]{0.3\linewidth}
   		\includegraphics[width=6.1cm, height=3.5cm]{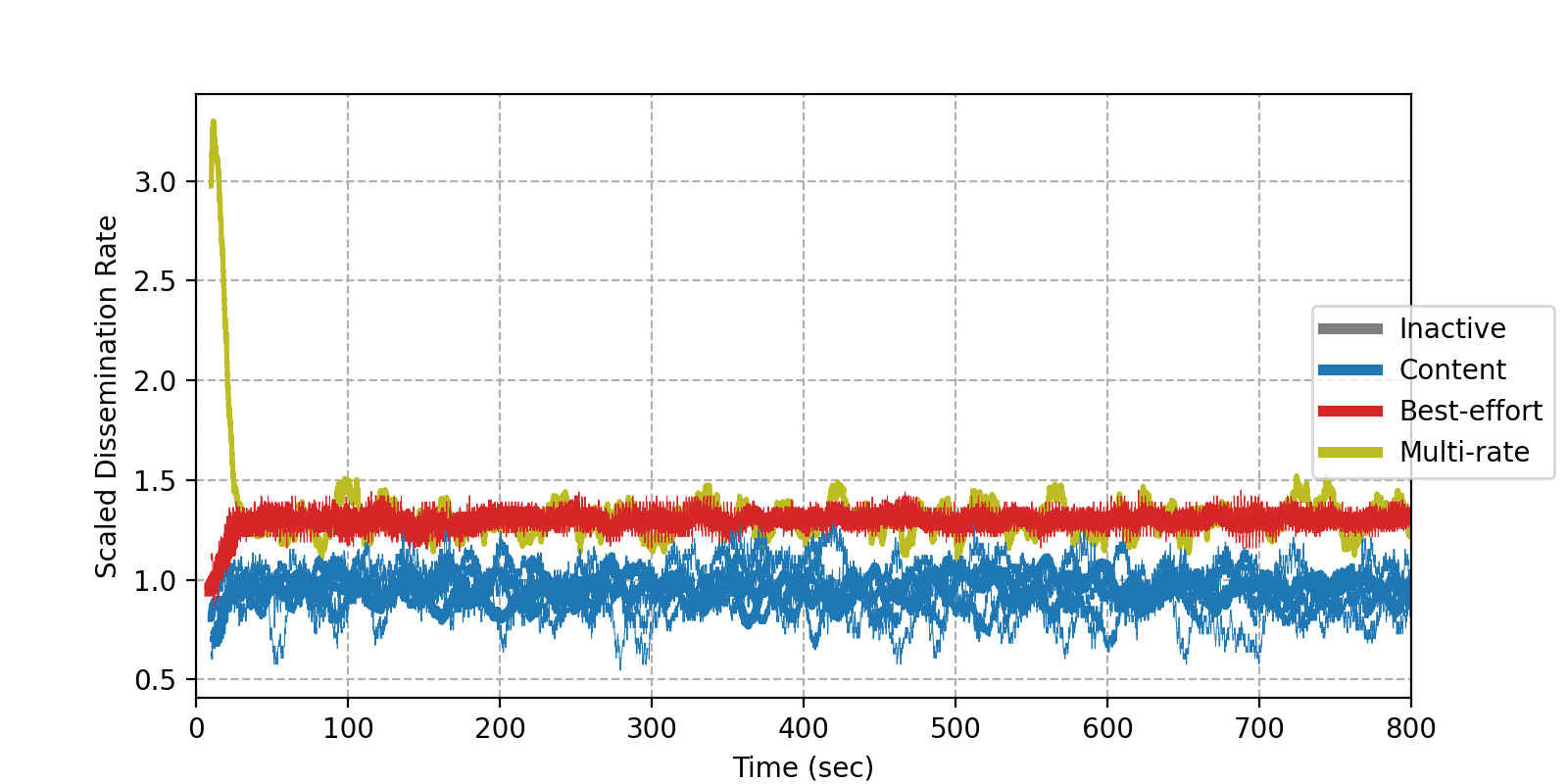}
    	\end{minipage}
	\label{fig:ScaSinMulDisRate}
    }
    	\hspace{1mm}
    	\subfigure[]{
		\begin{minipage}[b]{0.3\linewidth}
			\includegraphics[width=6.1cm, height=3.5cm]{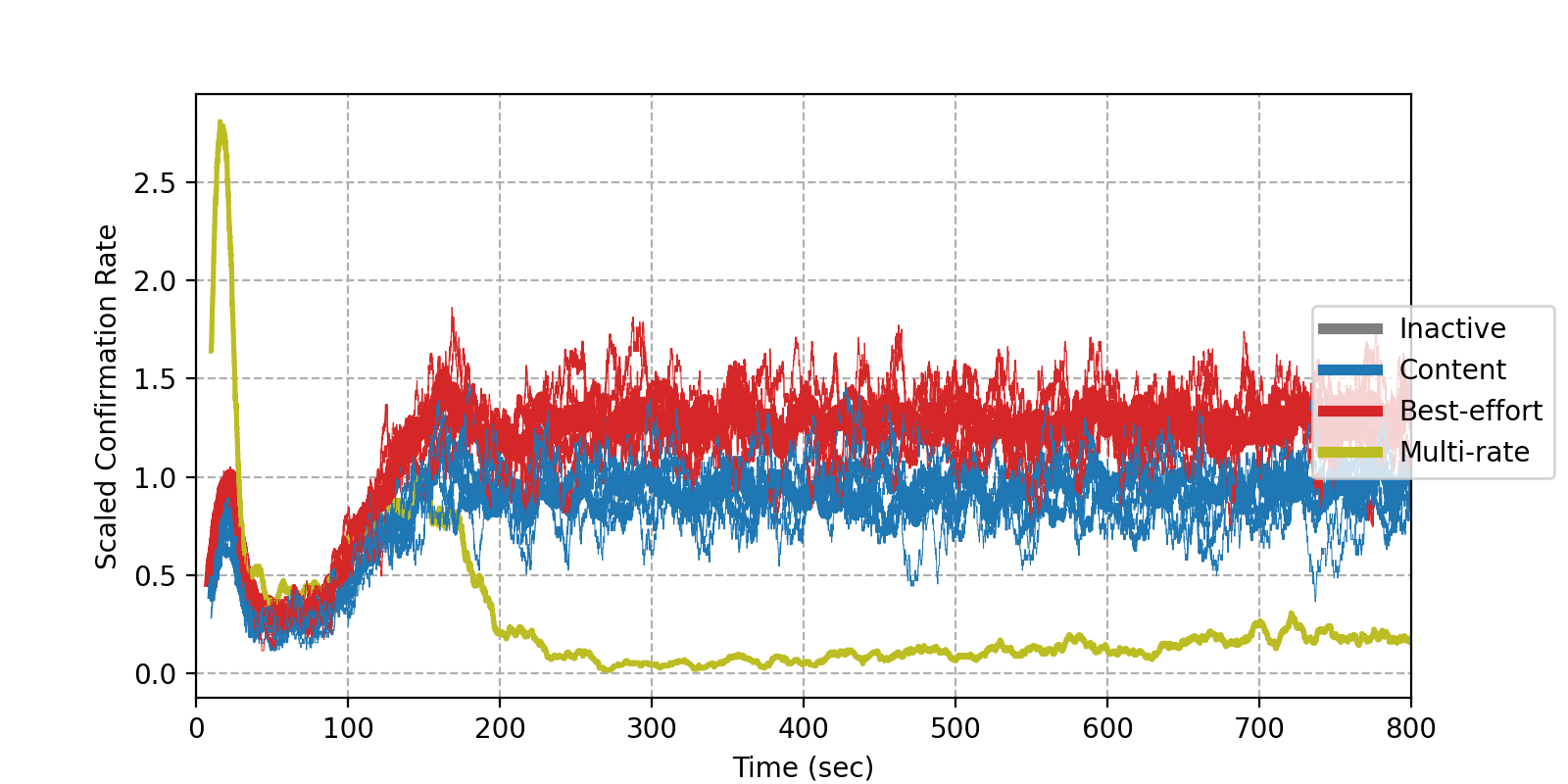}
		\end{minipage}
		\label{fig:ScaSinMulConfirRate}
}
    	\hspace{1mm}
    	\subfigure[]{
		\begin{minipage}[b]{0.3\linewidth}
			\includegraphics[width=6.1cm, height=3.5cm]{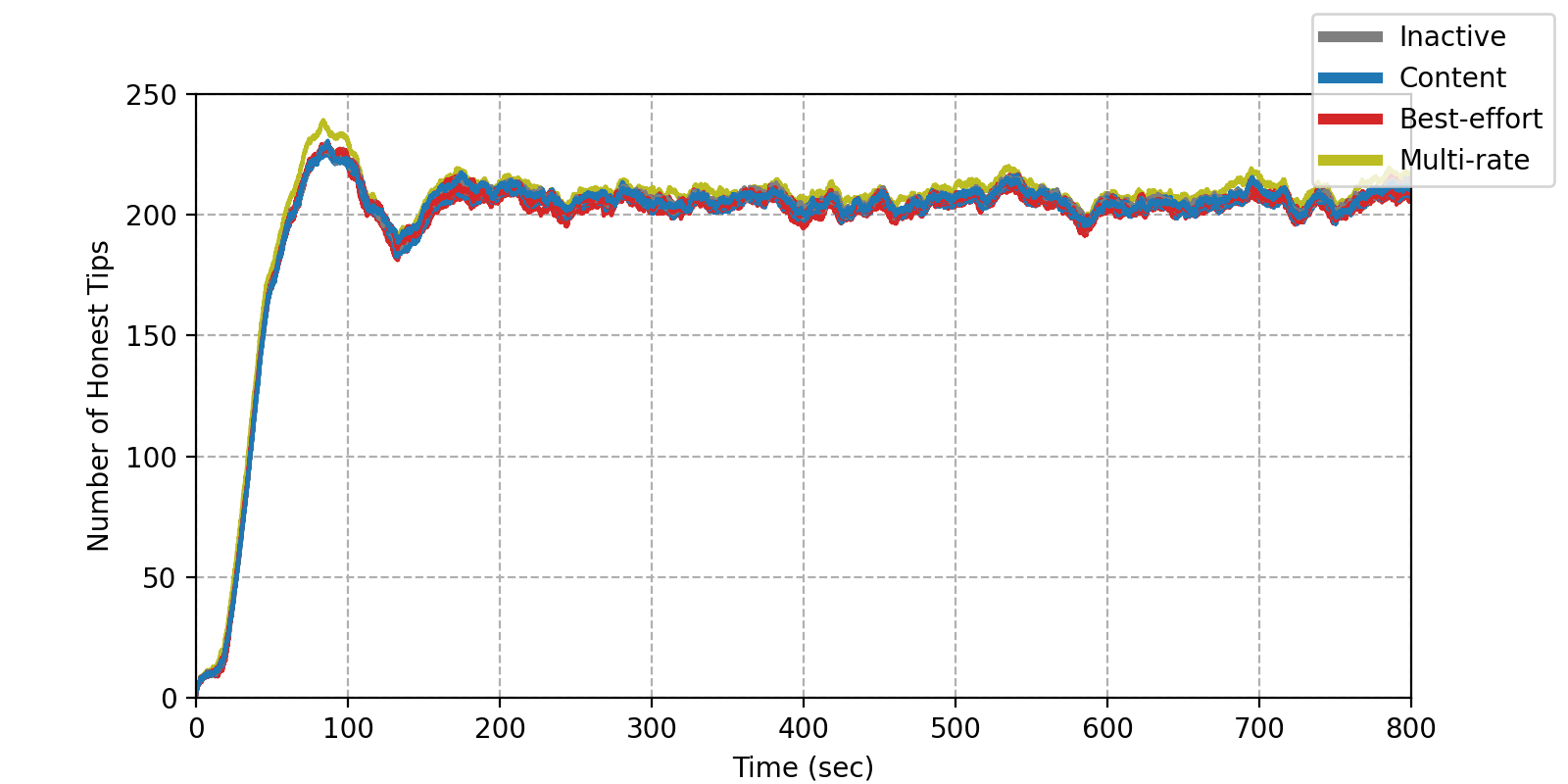}
		\end{minipage}
		\label{fig:Tips_MuR}
}
	\caption{$A_1$: Single spamming attacker (a): Scaled dissemination rate across all nodes.
(b): scaled confirmation rate across all nodes. (c): The number of tips from honest nodes;
$A_2$: Single multi-rate attacker (d): Scaled dissemination rate across all nodes.
(e): scaled confirmation rate across all nodes. (f): The number of tips from honest nodes. }
\label{fig:InboxMultip}
\end{figure*}
 The reputation-scaled dissemination rate is depicted in Figure \ref{fig:ScalSinSpaDisRate}. The colour of each line denotes the mode of the node to which it corresponds, as indicated in the legend, the thickness of each line is set to be proportional to its reputation. The scaled dissemination rate of best-effort and content nodes converge to a relevant constant value respectively. Hence, fair access to the network is ensured for each node. The reason that the malicious node also has a constant value is it can still issue blocks and attempt to forward its blocks to other nodes, although most of its blocks get dropped.

The reputation-scaled confirmation rate is depicted in Figure \ref{fig:ScaSinSpaConfirRate}, which shows that the confirmation rate of nodes is also proportional to the node's reputation for each mode of issuer. There are more confirmed blocks issued by high-reputation nodes because high-reputation nodes issue (or are allowed to issue) more blocks. The scaled confirmation rate values of best-effort and content nodes converge to constant values, while the confirmation rate for the malicious node is around 0, as most of its blocks are either dropped or get severely delayed and hence do not get selected as tips due to the PCT condition. As a result, these blocks do not gain cumulative weight to become confirmed.
 Figure \ref{fig:Tip_size} depicts the number of tips for honest nodes. As can be observed, the equilibrium point for all nodes' tip pool size is around $200$, even in the presence of attackers. This ensures that honest nodes get selected quickly by tip selection and do not experience significant delays in becoming confirmed.

\subsection{$A_2$: single multi-rate attacker}
Here we consider the attack $A_2$, in which a multi-rate attacker is in the network. The corresponding results for this attack scenario are shown in Figure \ref{fig:ScaSinMulDisRate}, which shows the scaled dissemination rates, Figure \ref{fig:ScaSinMulConfirRate} which shows the scaled confirmation rates, and Figure \ref{fig:Tips_MuR}, which shows the number of tips from honest nodes. It is clear by comparing these results to those in the previous subsection that our PCT condition is just as effective at dealing with a multi-rate attacker as it is for a spammer. The plots for $A_2$ are almost identical to those of $A_1$, so we will not repeat our analysis from the previous subsection.

\subsection{$A_3$: benchmark experiment (multi-rate attacker, no PCT)}

We evaluate and benchmark the effectiveness of our approach by repeating attack scenario $A_2$, but without the PCT condition. 
\begin{figure}[ht]
\centering
\includegraphics[width=0.95\columnwidth]{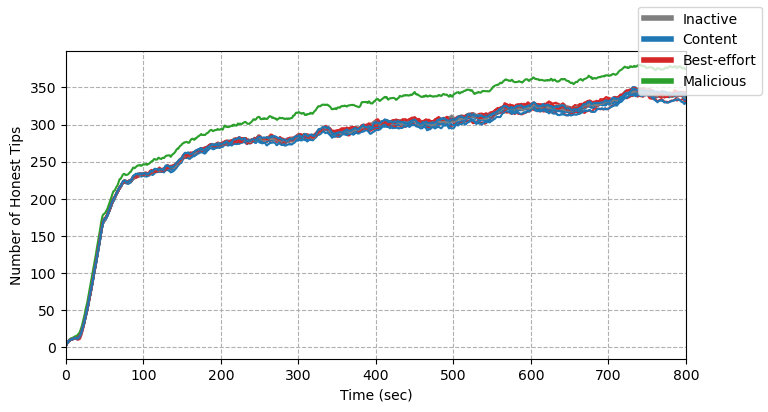}
\caption{ $A_3$: benchmark experiment: the number of tips from honest nodes.}
\label{fig: BenchTipSize}
\end{figure}
\begin{figure}[ht]
\centering
\includegraphics[width=0.98\columnwidth]{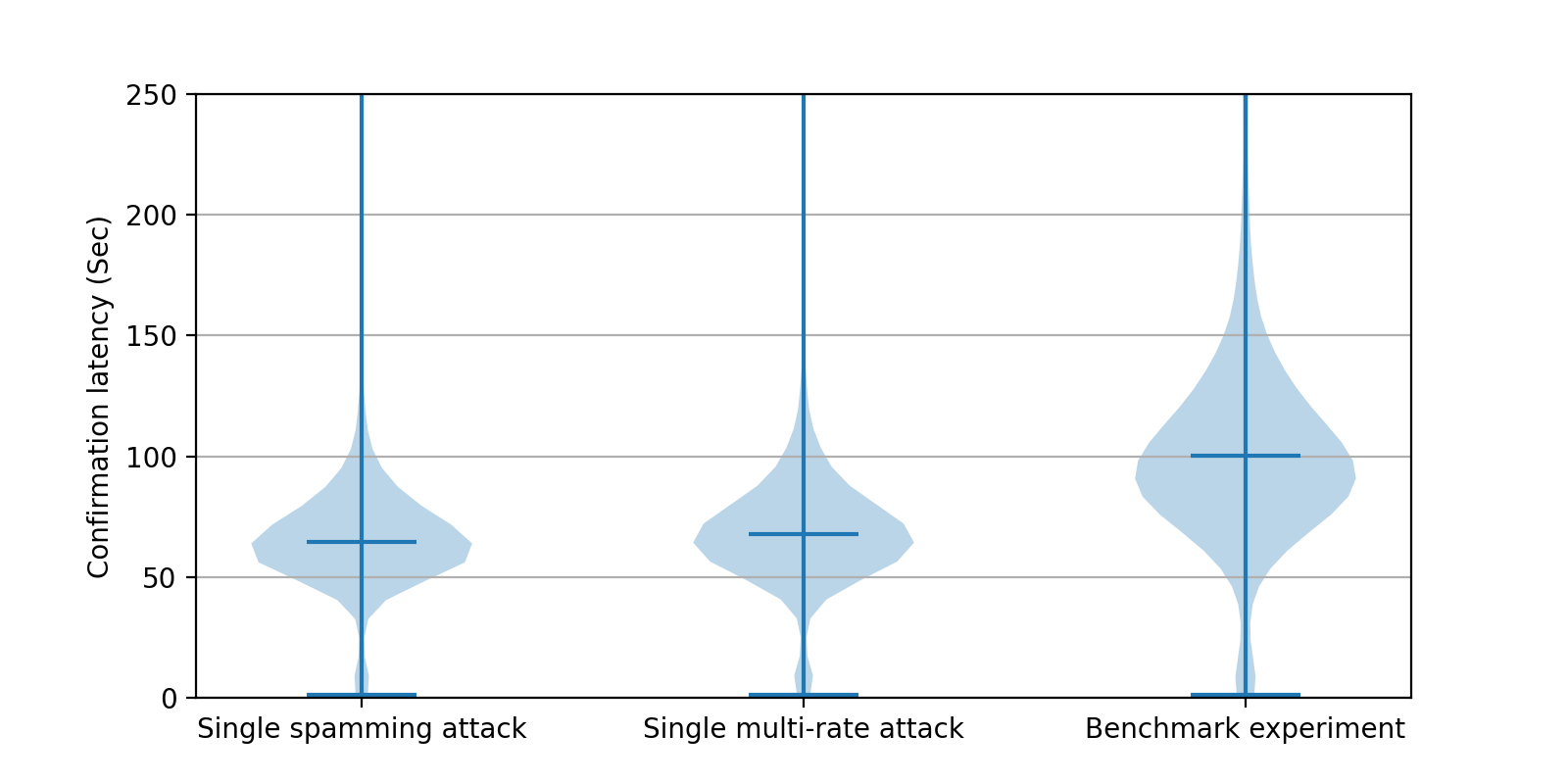}
\caption{The confirmation latency of all blocks in the network under different scenarios: single spamming attack, single multi-rate attack and benchmark experiment. }
\label{fig: VoConflatenNew}
\end{figure}
As illustrated in Figure \ref{fig: BenchTipSize}, when the PCT condition is not implemented, the number of tips increases continuously rather than converging to a steady state. With an extended simulation time, we can expect the tip pool size to continue to grow significantly. As a result, we can expect confirmation times to increase due to old tips being selected, further slowing the growth of cumulative weight for honest nodes.

Indeed, this expected behaviour is observed in Figure \ref{fig: VoConflatenNew}, which shows the distribution of confirmation latency for scenarios $A_1$, $A_2$ and the benchmark $A_3$. In $A_1$, the single spammer with PCT condition used, the average confirmation latency is approximately 64 seconds, and the maximum latency is around 110 seconds. In scenario $A_2$, the single multi-rate attacker with PCT condition used, the confirmation latency is slightly longer, with an average value of around 70 seconds and a maximum value of approximately 120 seconds. In the benchmark experiment, however, it is evident that in the absence of the PCT condition, the average and maximum confirmation latency increase substantially to around 100 seconds and 190 seconds, respectively.

\section{Conclusions and Future Work}
In this paper, we presented a past-cone confirmation time (PCT) condition for DAG-based distributed ledgers. The proposed algorithm ensures the security, robustness and consistency of the network even in the presence of attacks. Furthermore, simulations are provided to illustrate the efficacy of the proposal. As noted in Section~\ref{sec: simulations}, our results demonstrate steady-state behaviour and provide a comparative analysis, but the performance figures, such as confirmation rates and latencies, are highly dependent on parameter choices and are not indicative of real network performance. As such, future work should focus on verifying these results in real networks and obtaining performance figures. For example, IOTA's GoShimmer prototype\footnote{\url{https://github.com/iotaledger/goshimmer}} could be a suitable testbed for our proposal. Another focus of this future implementation work should be on quantifying the computational burden of the PCT check and improving the efficiency of the solution, if necessary. 

\bibliographystyle{unsrt}
\bibliography{sample}
\end{document}